\documentclass[12pt]{iopart}
\usepackage{times}
\usepackage{graphicx}

\begin{document}

\title{Tunneling properties of quantum dot arrays in strong
magnetic field}
\author{A.\ Aldea$^{1}$, V.\ Moldoveanu$^{1,2}$ and B.\ Tanatar$^{2}$}
\address{$^1$National Institute of  Materials Physics, POBox MG7,
Bucharest-Magurele, Romania}
\address{$^2$Department of Physics, Bilkent University, 06533 Ankara, Turkey}

\begin{abstract}
We develop a formalism suitable for the study of transport properties of
coherent multiple dots which captures and explains the  
experimentally observed features in terms of spectral properties 
of the system. The multiplet structure of the transmission spectrum
and the role  of the intra-dot and inter-dot Coulomb interaction 
are pointed out. The evolution of the sub-peaks with the magnetic 
field in the quantum Hall regime is shown.
We suggest a specific oscillatory behaviour of the Hall resistance in 
strong magnetic field
which can be experimentally tested.
\end{abstract}

1)
The electronic transport through coupled quantum dots became a 
topic of interest once the basic phenomena in single dots were 
satisfactorily understood. The Coulomb oscillations at vanishing 
magnetic field were investigated and the role of the inter-dot
coupling for the multiple peak structure of the conductance was
shown \cite{W95,W96}. 
The next step was accomplished by Livermore {\it et al}. \cite{Liv99} 
who measured the conductance through coupled dots
in the quantum Hall regime. It was shown that the conductance 
peaks undergo shifts and also modulation as the magnetic field 
is varied continuously. The quantum dot arrays are
promising systems of investigation because of the analogy with 
"artificial molecules" and possibility of physical realizations
of quantum bits (qubits).

We develop a theoretical approach of the tunneling problem 
through arrays of quantum dots in strong magnetic field with the aim
to elucidate several aspects: the mechanism of the multiplet formation,
the effects of the intra- and inter-dot Coulomb interaction and effects
of the strong magnetic field.
Our approach is based on a  non-Hermitian tight-binding Hamiltonian 
used in the Landauer-B\"uttiker formalism with many terminals \cite{MAMN01}. 
The non-Hermiticity of the Hamiltonian arises by
including  the contribution of the terminals  in the Hamiltonian of
the dot system in all powers of the perturbation theory.

We emphasize that 
in our approach the dot array is considered as a quantum-mechanical
coherent system. Each quantum dot is modelled as a 2D rectangular plaquette
and the coupling between them is ensured by tunneling terms (see Fig.1).
In what concerns the electron-electron interaction
we go beyond the usual {\it constant  capacitance model} and describe both the
intra- and inter-dot electron-electron interaction by a Coulomb-type  term.

\begin{figure}[htq]    
\begin{center}
      \includegraphics[angle=-90,width=8.0cm]{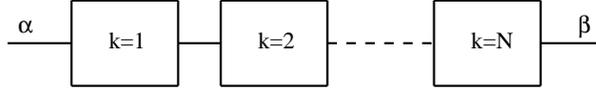}
\end{center}
\caption{The sketch of an array of $N$ quantum dots coupled to the leads
$\alpha$ and  $\beta$}
\label{desen}
\end{figure}
\vskip 0.4cm
2)
The formalism is presented shortly in what follows and represents the
generalization of the method which was used in Ref.4 for describing 
the transport properties of a single dot. 
For an array composed of $N$ dots, the Hamiltonian  contains the following  
terms:
\begin{equation}\label{1}
H^N=\sum_{k=1}^N H^k+\tau\sum_{k=1}^{N-1}H_{t}^{kk+1}
+U\sum_{k>k'}^NH_{ee}^{kk'}.
\end{equation} 
The index $k$ counts the dots in the array, $H^k$ is the Hamiltonian of an
individual dot  which includes the intra-dot electron-electron interaction,  
$H_{t}^{kk+1}$ describes the tunnel-coupling of 
consecutive dots, while $H_{ee}^{kk'}$ is the Coulomb interaction between
electrons located in different dots (i.e., the last term represents the 
so-called electrostatic inter-dot coupling).
The 2D spinless   Hamiltonian $H^k$ is considered  in the Hartree 
approximation , while the perpendicular magnetic field  is 
taken in the Peierls representation :

\begin{equation}
H^k=\sum_{i\in QD_k} 
\Big(V_{g}^{k}+U\sum_{j(\neq i)}{\langle n_{j}\rangle\over |j-i|}\Big)
c_i^{\dagger}c_{i}
+ t^{D}\sum_{<i,i'>}e^{i2\pi\phi_{ii'}}~c_i^{\dagger}c_{i'} \,.
\end{equation}
    
Here $c^{\dagger}_i(c_i)$ are the  creation (annihilation) 
operators  in localized states indexed by $i\in QD_k$ and $t^D$ is
the nearest-neighbour hopping integral in the dots.
The phase $\phi_{ii'}$ comes from the Peierls substitution and 
accounts for the magnetic flux through the
unit cell of the lattice measured in quantum flux units $\phi/\phi_0$. 

Now we have to built up the non-Hermitian Hamiltonian $H_{eff}$ which takes 
into account the coupling of the multiple dot system with the leads
carrying the external current; 
we note that $H_{eff}$  depends on the energy $E$ of the incident electron:
\begin{equation}\label{eff2}
H^N_{eff}(E)=H^{N}+\tau^2\sum_{\alpha}~e^{-ik}c^{\dagger}_
{\alpha}c_{\alpha}, ~~ E=2\cos{k} , 
\end{equation}
where the index $\alpha$ denotes the sites where the leads are stuck 
to the dots.

Finally, one has  to calculate the retarded Green
function $G^+(E)=(E-H^N_{eff}+i0)^{-1}$ which is to be used in the Landauer-
B\"uttiker formula in order to obtain the  conductance $g_{\alpha,\beta}$
\cite{MAMN01}.

Let us enumerate the parameters of the problem : 
i) the tunnel-coupling between the dot-system
and the leads described by the parameter $t_{LS}$, 
ii) the tunnel-coupling between dots in the array $\tau$,
iii) the strength of  the Coulomb interaction $U$ and  iv) the  gate voltage
$V_g^k$ which can be applied on any dot $k$.
The hopping integral  in the dots is taken as $t_{D}=1$, and
with this choice all the other energies $V_g$, $t_{LS}$ and $\tau$  
are measured in units of $t_{D}$.

For our two-dimensional  problem the calculation of $G^+$ has to be done 
numerically, and the
mean occupation number of each site $\langle n_i\rangle$ in Eq.\,(2) 
has to be calculated self-consistently.
The numerical effort depends on the number of dots in the array and on the 
dimension of each dot. When the electrostatic inter-dot coupling
is neglected,  
this effort can be much reduced by using an iterative procedure,
which allows the calculation of transmittance through any number of dots 
in the array, if the one-dot problem is solved.

We mention that the tight-binding model has already been used
for the study of transmittance of quantum dot arrays in the work of
Kirczenow \cite{Kir92} where each dot is associated with a single atom 
in the lattice, omitting thus the internal structure of the dots.
We perform here an important step forward 
by describing each dot as a finite two-dimensional plaquette.
\begin{figure}[htq]
\begin{center}
      \includegraphics[angle=-00,width=22.0cm]{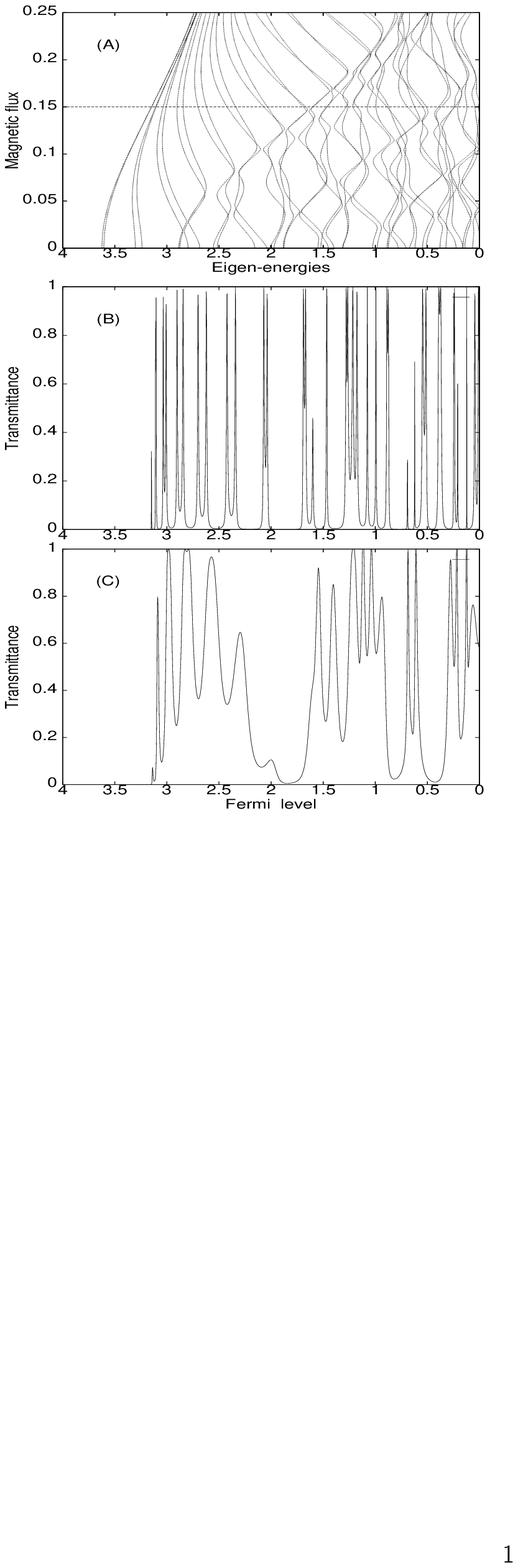}
\vspace*{-14cm}
\caption{ A) The degeneracy lifting of the eigenvalues for the Hofstadter-like
spectrum of a double dot at $\tau =0.4$.~
B) The corresponding  split peaks in the transmittance spectrum 
at magnetic flux $\phi/\phi_0=0.15$ and $t_{LS}=0.4$. 
C) The same for $t_{LS}=1.0$.
The system consists of two dots of dimension $5\times8$ sites each.}
\end{center}
\label{ox_spec}
\end{figure}

\vskip 0.4cm
3)
In small quantum dots, which is the topic of our interest,
the size quantization  compete with the charging effects produced by the
electron-electron interaction. So, we discuss first the
case of non-interacting electrons in order to distinguish afterwards 
the effects due to the interaction.
The  quantum system has a specific energy spectrum
depending on the number of dots in the array and
on the coupling constant between them which shows similarities to
the Hofstadter spectrum. However two main differences appear: the spectrum 
of the array exhibits a multiplet structure  due to the tunnel-coupling
between dots which lifts the degeneracy of the energy levels
and, on the other hand, the spectrum contains edge states which are
induced by the vanishing boundary conditions  
imposed to the wave function.

We note that in the case of many coupled dots the 
difference between
edge and bulk states is difficult to be realized in geometrical terms  so
that, essentially, they differ by their chirality. This is even more stringent 
in the presence of the  electron-electron interaction, when
geometrical deformations of the eigenstates occur.
Since the array is a coherent quantum system the eigenfunctions have the
character of  molecular states. 
When one of the dots
in the array is detuned (by applying a gate potential on it) the
molecular states are  scrambled with important effects 
on the longitudinal and transverse (Hall) conductances.

Evidently, the transmittance spectrum as a function of Fermi energy
($E_F$) has to correspond to the structure of the energy spectrum.
However, this depends on the strength of the coupling between 
the system of dots and the external leads, expressed by the  
parameter $t_{LS}$. More explicitly, a multiplet can be put 
into evidence only if the imaginary part of
$G^+(E_F)$ is less than the  interlevel distance and 
such a situation occurs only  for small  $t_{LS}$. 
For a double dot this is shown in Fig.2A and B where
the correspondence between the energy levels and the peaks of the
transmittance is evident indicating the resonance tunneling process
of the electron through the coherent double dot structure.
On the other hand, a  large
value of this parameter, as it is considered in Fig.2C, spoils the 
resonant aspect and the one-to-one correspondence between the
energy levels of the isolated quantum system and the transmittance peaks.
In this case, other aspects like the Fano profile of the transmittance peaks 
or the Fano zeros  are the topics of interest.

\vskip 0.4cm
4)
We now look for the effects of the interaction
on the transmittance spectrum of multiple dots.
From the  single dot problem we have learned that, at least in
the self-consistent Hartree approximation, the electron-electron 
interaction gives rise to an increase in the level spacing
and of the width of transmission peaks \cite{MAMN01}.
We consider that the same thing occurs
also for coherent multiple dots and Fig.3 confirms these 
expectations when compared with Fig.2B.
The same Fig.3 also shows the increase of the splitting with
increasing tunneling between dots and in fact Fig.3B indicates already
the saturation of the splitting effect. The saturation is non-linear in
the coupling parameter $\tau$ and 
can be put into evidence in a theoretical calculation only by
taking into account all order of the perturbation.
(For the sake of clarity of the drawings we show the results for $ N=2$).

The results shown in Fig.3 are obtained by taking into account the  
total Coulomb interaction. However, one asks
usually how important the inter-dot interaction is.
This contribution represents an additional term compared to the
intra-dot electron-electron interaction, and it is 
generally accepted that the strength $U$ of the inter-dot interaction 
is smaller than in the intra-dot case
because of the screening produced by the metallic gates which exist 
between neighbouring  dots \cite{Usmall}. 

The inter-dot term should accentuate the interaction effects  
known for the transmission through a single dot.
Furthermore, since the interaction energy is positive, in the presence of
the additional inter-dot coupling the whole transmittance spectrum should
be pushed upwards on the energy scale, as it can be noticed by
comparing Fig.4A and B. However, the comparison indicates a rather
messy behaviour in the domain $E_F>0$; this is because here the 
spectrum is occupied mainly by bulk states which are more sensitive to
the interaction, meaning that they
undergo easily  geometrical deformations and ,in addition,
edge states are intercalated among them \cite{cond-mat02}.
\begin{figure}[htq]
\begin{center}
\vspace*{-1cm}
\hspace*{-1cm}
      \includegraphics[angle=-00,width=18.0cm]{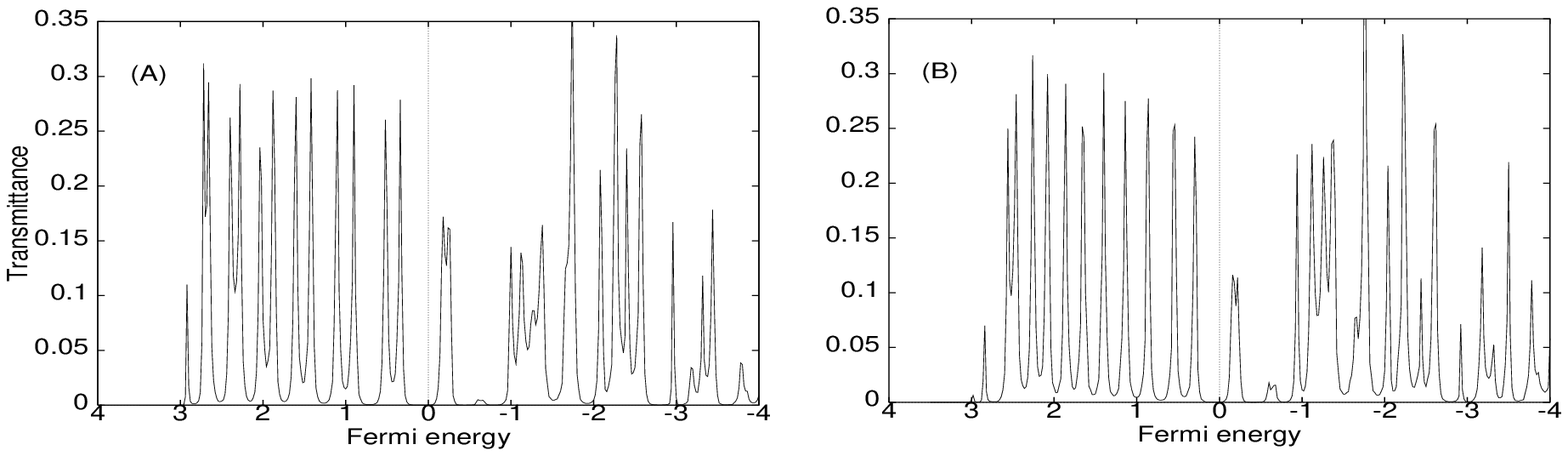}
\vspace*{-17.5cm}
\caption{ The dependence of the transmittance spectrum 
of a double dot on the tunnel-coupling 
between dots A) $\tau=0.4$, B) $\tau=0.8$.}
\vspace*{-1cm}
      \includegraphics[angle=-00,width=20.0cm]{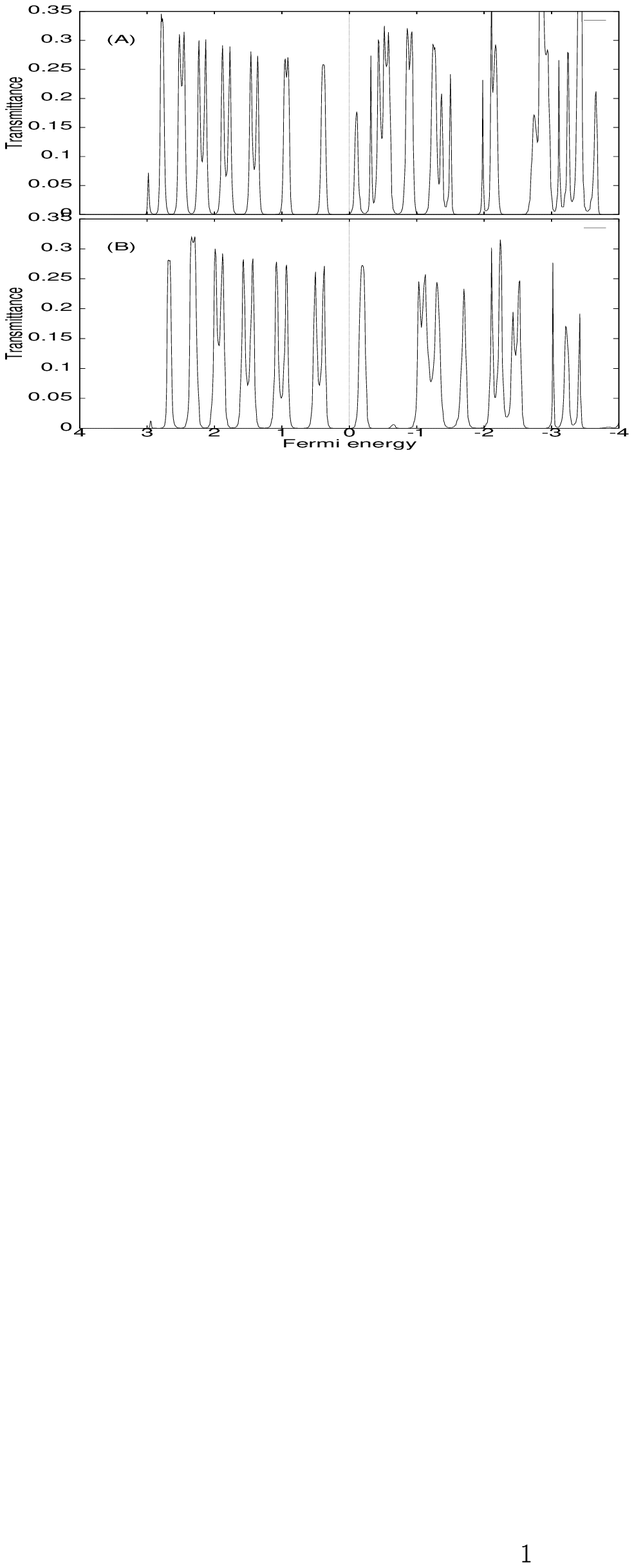}
\vspace*{-18cm}
\caption{Contribution of the inter-dot Coulomb interaction:
A) Transmittance spectrum in the presence of intra-dot interaction,only.
B) The same  in the presence of both intra-dot and inter-dot 
electron-electron interaction. The similarity is lost in the region 
$V_g>0$ filled mainly with bulk states
which are more sensitive to the additional interaction term.($U=0.5,
\tau=0.2, t_{LS}=0.4$.}
\end{center}
\label{ox_trans}
\end{figure}
\vskip 0.4cm
5)
Our approach helps in understanding the recent
experiments  by Livermore et al \cite {Liv99}. The Fig.3a-c in  
Ref.3 reveal that, in strong magnetic field, the transmittance of 
a double dot behaves as follows: i) at a given magnetic field but
increasing inter-dot coupling, one starts with single peaks (figure a), 
then twin peaks appear (figure b) and finally, at large inter-dot 
coupling,
the saturation consisting again in individual peaks occurs (figure c).
ii) with increasing  magnetic field 
the position of each peak shifts linearly (in spite 
of a zig-zag appearance) versus larger gate voltages. Both these 
features can be
described by our model even  without  considering the interaction.
Our results shown in Fig.5 indicate a striking similarity to the 
experimental situation. They indicate both the increase of the
distance between the twin peaks with the increasing inter-dot tunneling 
$\tau$ and the drift of the peaks at the variation of the magnetic field.
The blurred, quasi-zig-zag behaviour of this drift is not however much 
evident in our 
calculations which are performed for non-interacting electrons. So, the
idea expressed in Ref.3  that this behaviour is due to interaction effects 
turns out to be correct;
our calculations for the interacting case are in progress.
Our approach puts however into evidence  that such a  regular 
behaviour, as observed in the experiment, cannot occur in any range of 
the energy spectrum, but only if the
Fermi level lies in a region covered by edge states.
Fig.2A shows also that only the edge states are
regularly distributed and well separated on the energy scale.

\begin{figure}
\begin{center}
\vspace*{-2cm}
\hspace*{1cm}
      \includegraphics[angle=00,width=16cm]{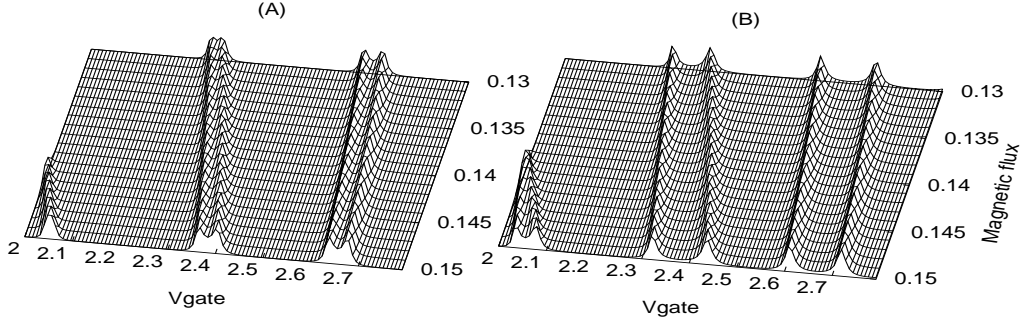}
\end{center}
\vskip -14cm
\caption{The drift with the magnetic field of the twin peaks from the
conductance spectrum of a double dot for two values of 
the tunnel-coupling :
A) $\tau=0.2$,~~B) $\tau=0.6$.}
\label{ox_liver1}
\end{figure}

\vskip 0.4cm
6)
The properties of the transmittance matrix impose specific 
behaviour to the Hall resistance.
In the presence of a constriction between the mesoscopic system
and the  external leads,
even in strong magnetic field, the Hall resistance
$R_H$ exhibits oscillations instead of the usual quantum Hall plateaus.
Since any minimum in transmittance give rise to a maximum in $R_H$,
the  minima of the multiple dot transmittance  produced by  splitting 
result in  small maxima of the Hall resistance located in-between
two large maxima. For a double dot with the configuration of leads shown in the 
sketch below (Fig.\,6) the Hall resistance $R_H$ is given in terms
of the conductance matrix elements  by the following 
expression :
\begin{equation}
R_{H}=(g_{21}g_{43}-g_{23}g_{41}-
g_{12}g_{34}+g_{32}g_{14})/2D ,
\end{equation} 
where $D$ is a positive $3\times 3$ subdeterminant of
the $4\times4$ matrix $g_{\alpha\beta}$
and the  result is shown in Fig.\,6.  
We have found that as the
saturation is installed all maxima of $R_H$ become equivalent.
To the best of our knowledge such a property was not yet observed 
experimentally.
\hskip -3cm
\begin{figure}[htq]
\begin{center}
\hspace*{3.cm}
      \includegraphics[angle=-90, width=14.0cm]{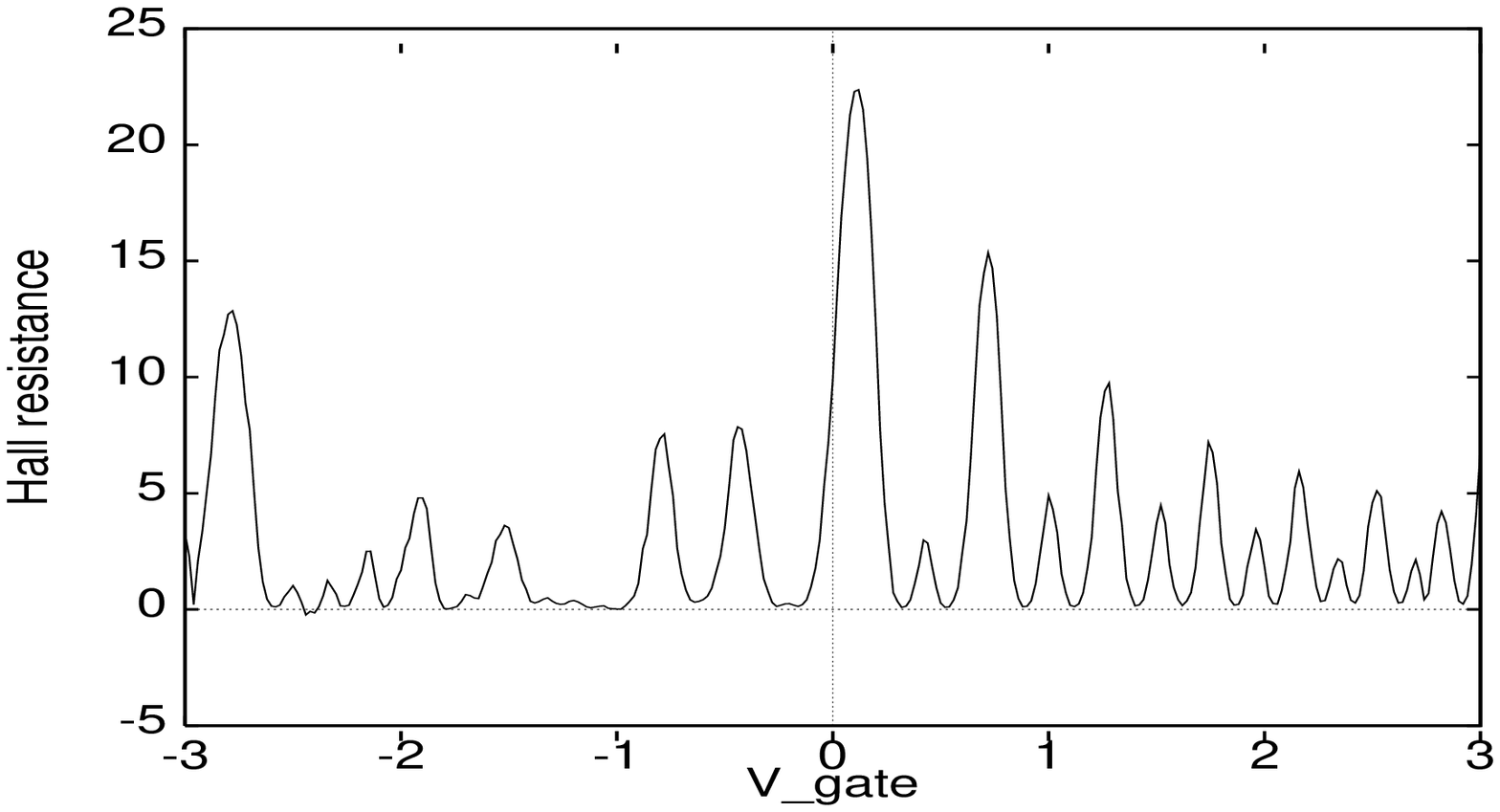}
\vskip -4.5cm
      \includegraphics[angle=-90, width=4.0cm]{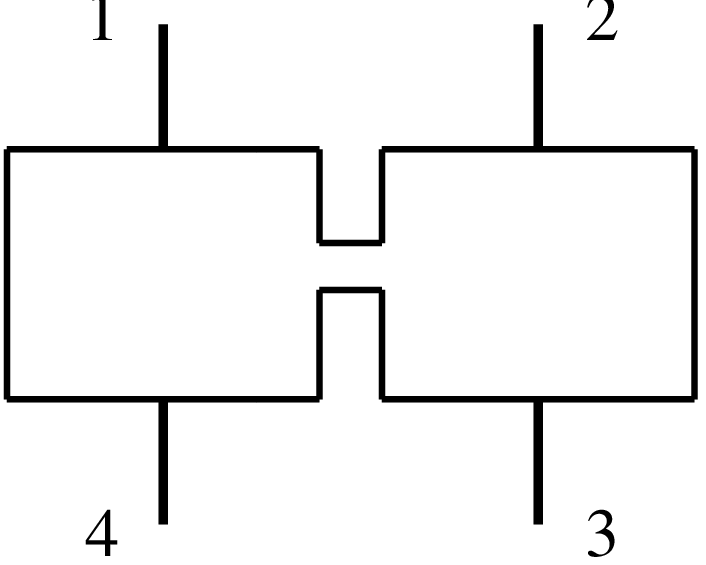}
\end{center}
\caption {The oscillatory behaviour of the Hall resistance for
a double dot in  high magnetic field regime ($\phi/\phi_{0}=0.15 $)
and in the presence of a 
dot-lead constriction ($t_{LS}=0.5$); the inter-dot tunnel-coupling 
is $\tau=0.3$.
The small amplitude oscillations due to the splitting are more evident 
in the range $V_g>0$ where the spectrum is covered by edge states.
The same gate potential is applied on both dots.}
\label{ox_hall}
\end{figure}

\vskip 40cm
In conclusion, an appropriate form of the  Landauer-B\"uttiker approach has
been used  to describe the electronic transport through an array of coupled 
quantum dots placed in a strong perpendicular magnetic field. Within the
tight-binding model and the self-consistent Hartree approximation
for both intra- and inter-dot Coulomb interaction, 
we have shown that the  multiplet structure is determined by the
tunnel-coupling between dots, while the Coulomb interaction gives rise to
a significant width of the peaks and increase of the splitting.
The peak splitting shows saturation at the perfect tunnel-coupling of the dots.
The quantum states of the edge-type are much more robust to the 
electron-electron interaction then the bulk-type states which undergo
strong deformations with consequences for the transmittance spectrum.
We suggest that the regular drift of the sub-peaks with the variation 
of the magnetic field, which has been observed experimentally \cite{Liv99},
occurs only if the Fermi level lies in a domain of the energy spectrum
filled with edge-states.
The multiplet structure of the transmittance yields small oscillations 
of the Hall resistance in the quantum  regime.

Further interesting problems such as the magnetic drift in the 
presence of the electron-electron interaction, the detuning effects 
on the Hall resistance, and the coherence properties of the array 
(as given by the phase of the transmittance) will be the subjects
of future work which is under way.

{\bf Acknowledgments.} This work was supported by Grant CNCSIS/2002
and Romanian Programme for Fundametal Reserch and
by NATO-PC TUBITAK Programme.

\end{document}